\documentclass{aadebug}

\corr{0309027}{261}

\begin{document}

%%%%%%%%%%%%%%%%%%%%%%%%%%%%%%%%%%%%%%%%%%%%%%%%%%%%%%%%%%%%%%%%%%%%%%%%
\runningheads{Joel Huselius, Henrik Thane and Daniel Sundmark.}{Availability Guarantee for Deterministic Replay Starting Points in Real-Time Systems}

\title{Availability Guarantee for Deterministic Replay \\
Starting Points in Real-Time Systems\extranum{2}}

\author{
Joel~Huselius\addressnum{1}\comma\extranum{1},
Henrik~Thane\addressnum{1},
Daniel~Sundmark\addressnum{1}
}

\address{1}{
Department of Computer Science and Engineering,
M\"{a}lardalens University, 
P.O.Box 883,
SE-721 23 V\"{a}ster\aa{}s, Sweden
}
\extra{1}{E-mail:joel.huselius@mdh.se}
\extra{2}{This work was supported by the Swedish Foundation for
Strategic Research (SSF) via the research programme SAVE, the Swedish
Institute of Computer Science (SICS), and M\"{a}lardalen University.}

% This information will show up in `Document Properties' in Acrobat Reader
\pdfinfo{
/Title (Availability Guarantee for Deterministic Replay Starting Points in Real-Time Systems)
/Author (Joel Huselius et al.)
}

%%%%%%%%%%%%%%%%%%%%%%%%%%%%%%%%%%%%%%%%%%%%%%%%%%%%%%%%%%%%%%%%%%%%%%%%
\begin{abstract}
Cyclic debugging requires repeatable executions. As non-deterministic
or real-time systems typically do not have the potential to provide
this, special methods are required. One such method is replay, a
process that requires monitoring of a running system and logging of
the data produced by that monitoring. We shall discuss the process of
preparing the replay, a part of the process that has not been very
well described before.
\end{abstract}


\section{Introduction\label{introduction}}

{\it Cyclic debugging} is the commonly used term for the process of
debugging a system using an ordinary debugger (e.g., gdb). That
process normally restarts the system repeatedly (with the same input)
to pinpoint the bug, hence ``cyclic''. This method of debugging relies
on that the same execution can be deterministically recreated at
command over and over again. Replay has been proposed to realize
cyclic debugging of systems that do not fulfill this requirement,
systems that incorporate elements of non-deterministic behavior and/or
time-dependence \cite{huselius+_2003, stewart_gentleman_1997,
zambonelli_netzer_1999} (e.g. real-time systems).

Replay can be described as creating a facsimile of an execution based
on a previous recording. The general idea behind replay is to, by
inserting {\it probes} \cite{huselius_2002} into the system, {\it
record} (to {\it monitor} and {\it log}) sufficient information about
a {\it reference execution} of the non-deterministic system to
facilitate the reproduction of a {\it replay} execution. The
information logged consists of {\it events} describing the execution
\cite{thane+_2003}: control-flow events (describing context-switches,
exceptions, and interrupts), and data-flow events (describing
checkpoints of task-states and input from the environment or from
other tasks). As an event is monitored, an {\it entry} with
information describing the event is logged into a {\it record} for
post-mortem usage.

Here, we assume {\it deterministic replay} \cite{thane+_2003}, which
does not assume that the log from the reference execution describes
the reference execution in its entirety; some sequences of the log may
be discarded before completion of the reference execution. This
interrupted coverage of the reference execution is a corollary of that
the space for storing records on is not infinitely large; thus, some
records may have to be discarded in favor of newer ones. In this
paper, we are concerned with the {\it eviction scheduler} that
controls the contents of the log.

This paper is a continuation on previous efforts \cite{huselius+_2003,
huselius_2002b}, where we presented a method for starting a replay
execution from {\it starting points} identified in the task code, and
concluded that successful management by the eviction scheduler of the
memory pool is vital to the performance of the replay execution. We
noted that the structure of the task code may be such that several
starting points exists, some of which may be unreachable if we cannot
guide the execution to them during the setup of the replay
execution. However, guiding the execution requires logged data that
describes that transition, the eviction scheduler must be responsible
for keeping required entries in the log in order to guarantee the
replay execution.

The previous work on eviction schedulers is not substantial, as with
this paper, we elevate the issue in the particular case of real-time
systems, and provide a method that is general, provided that the known
and thoroughly defined conditions listed in
\cite{huselius+_2003} are met. The paper is organized as follows:
Section \ref{related_work} recapitulates some previous work in the
area, Section \ref{ecetes} presents our method, and Section
\ref{conclusions} concludes the paper.

\section{Related Work\label{related_work}}

Stewart and Gentleman \cite{stewart_gentleman_1997} report that many
replay-solutions has made use of FIFO-queues for logging data from the
monitoring effort. Implementations vary from the
one-queue-for-all-entries Global FIFO (GFIFO), to
one-queue-per-source-of-entries Local FIFO (LFIFO). However, with
GFIFO, some old records that are essential to the replay may be lost
\cite{huselius+_2003}, and LFIFO requires the possibility to dimension
the relative sizes of the different queues.

Zambonelli and Netzer \cite{zambonelli_netzer_1999} proposed a method
that, by taking on-line decisions on whether to log or not to log a
monitored event, deviates from the strict FIFO-solution. However,
sometimes logging will debit the system with a jitter in the execution
time, and so will also the algorithm it self. As larger jitter will
force more extensive efforts for validation \cite{thane_hansson_2001},
an increase in jitter is counterproductive to the validation effort.

\section{The extended constant execution time eviction scheduler ECETES\label{ecetes}}

We have, in a previous publication \cite{huselius_2002b}, presented a
method called the Constant Execution Time Eviction Scheduler
(CETES). Similarly to the method proposed by Zambonelli and Netzer,
CETES took on-line decisions on how to organize the logging
efforts. Unlike the proposition of Zambonelli and Netzer, our method
had a constant execution time and logged all monitored events. The
main drawback of CETES was the requirement that all entries be the
same size. In this work, we propose the Extended Constant Execution
Time Eviction Scheduler (ECETES).

\subsection{Implementation}

ECETES allows a user to specify a set of queues where data can be
stored, the queues share a pool of memory, divided into atomic
records, where entries can be logged concurrently in one or more
records. The setup of the ECETES system requires the input-parameters:
size and location of the memory pool, the size of one record, the
maximum number of records per entry, and the number of queues.

As posted above, execution time jitter forces more extensive
validation efforts. In this setting, as such an increase would be
counter productive, we require that ECETES has no execution time
jitter. The functionality of the ECETES is very influenced by this
restriction on the implementation; given the same input-parameters,
the execution time of the implementation should be deterministic. When
a call is made to insert a new entry in the ECETES structure, the
required input-parameters are: The location of the queue in which to
store the entry, the location of the entry, the number of records
required to store the entry, the type of the record, and the current
time. The type of the record specifies if it is a control- or
data-flow entry, and the specific type of flow-entry.

We have decided upon the following solution: A call to insert a new
entry that requires $l$ records will cause ECETES to inspect one entry
of each queue, the $(l+1)$:th record counted from the end of the
queue. The inspected records are compared with respect to their age
and the properties of the queue to which they belong. At the end of
the comparison, a queue has been chosen that will suffer least if $l$
records are removed from it. From this queue, $l$ records are then
removed from the queue, and they are instead inserted into the queue
described by the input-parameters and can there accommodate the
requested entry. Before exiting, some internal structures of the
ECETES are updated.

The above mentioned queue-properties can be used to control the
operation of ECETES without modifying its execution time
characteristics. Currently, the implementation supports a similar set
of queue properties as did CETES: Minimum Temporal Span (MTL) that can
prevent young records from being evicted, Minimum Spatial Span (MSL)
that ensures the availability of a specified number of records in the
queue, Queue Priority (QP) relates queues with respect to their
importance (entries are evicted from low-priority queues provided that
no other constraint is violated).

\subsection{Using ECETES}

As noted in our previous work \cite{huselius+_2003}, {\it potential
starting points} for replay can be identified in the task code. A
potential starting point is a {\it starting point} if there is a
checkpoint of the task-state and a control-flow entry from that point
available in the log. In order to start a replay from a particular
starting point, it is required that the execution up on till the first
encounter of that checkpoint can be deterministic
\cite{huselius+_2003}. It is potentially required that a replay must
be used to guide the replay from one starting point to a consecutive
one in order to fulfill the stipulated requirement.

When using ECETES for recording, the following setup is intended: One
queue should collect all the control-flow of the system, separate
queues should store data-flow entries for each task. Task-constructs
that have more then one starting point are allocated one queue per
starting point in order to guarantee replay (in compliance with the
discussion in \cite{huselius+_2003}). The MSL of each queue is set to
guarantee the availability of at least on complete checkpoint. The
effectiveness of the solution is increased if a record size could be
established so that there are few records for each entry, and so that
there is not much redundant space lost due to incompatible entry
sizes.

\subsection{When is ECETES better then the alternatives?}

The only previously known method that fulfills all requirements posted
on an eviction scheduler is the LFIFO algorithm. As queue-sizes are
static, LFIFO has the potential drawback that queues must be
dimensioned pre-runtime.

We have still to perform the validation of ECETES, but we have hopes
that it should outperform LFIFO in situation where: the taskset has a
high degree of sporadicity, or where data-flow entries describing
checkpoints may come from different program counter values in the same
task (as described above, a task may have several starting points).

\section{Conclusions\label{conclusions}}

We have presented a new method called ECETES for managing the memory
available for logging, a new {\it eviction scheduler}. We have
described the situations where we expect ECETES to perform better then
traditional methods, but the validation is yet to be performed.

\bibliography{huselius_aadebug03}
\end{document}